\documentclass[prx,twocolumn,showpacs,superscriptaddress,preprintnumbers,floatfix,amsmath,amssymb]{revtex4-1}

\usepackage{amssymb}
\usepackage{graphicx}
\usepackage{dcolumn}
\usepackage{bm}
\usepackage{color}
\begin{document}

\title{Topological Dirac states in a layered telluride TaPdTe$_5$ with quasi-one-dimensional PdTe$_2$ chains}

\author{Wen-He Jiao}
\email[]{whjiao@zust.edu.cn}
\affiliation{Department of Applied Physics, Zhejiang University of Science and Technology, Hangzhou 310023, China}

\author{Xiao-Meng Xie}
\affiliation{Department of Applied Physics, Zhejiang University of Science and Technology, Hangzhou 310023, China}

\author{Yi Liu}
\affiliation{Department of Applied Physics, Zhejiang University of Technology, Hangzhou
310023, China}

\author{Xiaofeng Xu}
\affiliation{Department of Applied Physics, Zhejiang University of Technology, Hangzhou 310023, China}
\email[]{xiaofeng.xu@cslg.edu.cn}

\author{Bin Li}
\affiliation{New Energy Technology Engineering Laboratory of Jiangsu Province and School of Science, Nanjing University of Posts and Telecommunications, Nanjing 210023, China}
\affiliation{National Laboratory of Solid State Microstructures, Nanjing University, Nanjing 210093, China}

\author{Chun-Qiang Xu}
\affiliation{School of Physics and Key Laboratory of MEMS of the Ministry of Education, Southeast University, Nanjing 211189, China}

\author{Ji-Yong Liu}
\affiliation{Department of Chemistry, Zhejiang University, Hangzhou 310027, China}

\author{Wei Zhou}
\affiliation{Department of Physics, Changshu Institute of Technology, Changshu 215500, China}

\author{Yu-Ke Li}
\affiliation{Department of Physics, Hangzhou Normal University, Hangzhou 310036, China}

\author{Hai-Yang Yang}
\affiliation{Department of Physics, Hangzhou Normal University, Hangzhou 310036, China}

\author{Shan Jiang}
\affiliation{Wuhan National High Magnetic Field Center and School of Physics, Huazhong University of Science and Technology, Wuhan 430074, China}

\author{Yongkang Luo}
\affiliation{Wuhan National High Magnetic Field Center and School of Physics, Huazhong University of Science and Technology, Wuhan 430074, China}

\author{Zeng-Wei Zhu}
\affiliation{Wuhan National High Magnetic Field Center and School of Physics, Huazhong University of Science and Technology, Wuhan 430074, China}

\author{Guang-Han Cao}
\affiliation{Department of Physics, Zhejiang Province Key Laboratory of Quantum Technology and Devices, Interdisciplinary Center for Quantum Information, and State Key Lab of Silicon Materials, Zhejiang University, Hangzhou 310027, China} \affiliation{Collaborative Innovation Centre of Advanced Microstructures, Nanjing 210093, China}

\date{\today}

\begin{abstract}
We report the synthesis and systematic studies of a new layered ternary telluride TaPdTe$_5$ with quasi-one-dimensional PdTe$_2$ chains. This compound crystalizes in a layered orthorhombic structure with space group $Cmcm$. Analysis of its curved field-dependent Hall resistivity, using the two-band model, indicates the hole-dominated transport with a high mobility $\mu_{h}$ = 2.38 $\times$ 10$^3$ cm$^2$ V$^{-1}$ s$^{-1}$ at low temperatures. The in-plane magnetoresistance (MR) displays significant anisotropy with field applied along the crystallographic $b$ axis. The MR with the current applied along the $c$-axis is also measured in high magnetic fields up to 51.7 T. Remarkably, it follows a power-law dependence and reaches (9.5 $\times$ 10$^3$)\% at 2.1 K without any signature of saturation. The De Haas-van Alphen oscillations show a small Fermi-surface pocket with a nontrivial Berry phase. The Shubnikov-de Haas (SdH) oscillations are detected at low temperatures and under magnetic fields above 28.5 T. Two effective masses $m^*$ (0.26$m_e$ and 0.41$m_e$) are extracted from the oscillatory SdH data. Our first-principles calculations unveil a topological Dirac cone in its surface states, and, in particular, the topological index indicates that TaPdTe$_5$ is a topologically nontrivial material.
\end{abstract}

\maketitle

\section{Introduction}
Topological insulators (TIs) are characterized by robust gapless edge or surface states with a linear Dirac dispersion, which are a direct physical consequence of the nontrivial topology of the bulk band structure and protected by time-reversal symmetry \cite{Kane,Zhang}. Since the remarkable discovery of TIs, the search for new topological materials has been one of the most interesting topics in condensed matter physics. The classification of materials based on symmetry and topology has been extended to metals, semimetals and even superconductors. For example, three-dimensional (3D) topological Dirac and Weyl semimetals have conduction and valence bands dispersing linearly through nodal points \cite{Wan,Wang,Dai,Chen1,Chen2,Chen3,Xu,Huang}. When these corresponding bands are doubly degenerate, the materials are named as Dirac semimetals (DSMs) \cite{Wang,Dai,Chen1,Chen2}. In the presence of broken time-reversal symmetry or space-inversion symmetry, DSMs evolve into Weyl semimetals (WSMs), and each Dirac cone in DSMs splits into a pair of Weyl cones with opposite chirality due to the lifted spin degeneracy \cite{Chen3,Xu,Huang}. Lorentz invariance breaking can further give rise to type-II Dirac fermions with a tilted Dirac cone in DSMs \cite{Noh,Fei}, and type-II Weyl fermions with a tilted Weyl cone in WSMs \cite{Bernevig}. If Dirac band crossings take place along a one-dimensional line or loop, in contrast with discrete nodal points in conventional DSMs or WSMs, nodal points turn into the lines of Dirac nodes, which is the band feature of unconventional topological nodal-line semimetals \cite{Bian,Mao}. The topological materials usually display exotic properties such as large magnetoresistance (MR) \cite{Ong}, high charge carrier mobility \cite{Ong,Zhao}, chiral anomaly \cite{Chen,Jia}, exotic superconductivity \cite{Alidoust,Li}, and novel quantum oscillations \cite{Brien}, providing a platform to promote the wide-ranging applications in next-generation spintronics and quantum computing \cite{Kane,Zhang}.

Among the reported topological materials so far, the transition-metal tellurides take up a large proportionality. For instance, the binary tellurides Bi$_2$Te$_3$ \cite{Zhang1,Shen}, Sb$_2$Te$_3$ \cite{Zhang1} were proposed and evidenced to be TIs. (W,Mo)Te$_2$ \cite{Ali,Balicas} and $TM$Te$_2$ ($TM$ = Pd, Pt, Ni) \cite{PdTe2,PtTe2,NiTe2} were confirmed to host type-II Weyl and type-II Dirac fermions, respectively. Very recently, the ternary tellurides $M$$TM$Te$_4$ ($M$ = Nb, Ta; $TM$ = Ir, Rh) were theoretically predicted as a new series of type-II WSMs, and expected to host a minimum of four Weyl points within the first Brillouin zone by including spin-orbit coupling (SOC) \cite{TaIrTe4-1,MMTe4}. Experimental results verify the existence of Weyl points in TaIrTe$_4$ and NbIrTe$_4$ \cite{TaIrTe4-2,NbIrTe4-1,NbIrTe4-2}. Unconventional surface superconductivity, exhibiting quasi-one-dimensional (quasi-1D) and topologically non-trivial characteristics, was even observed in TaIrTe$_4$ \cite{TaIrTe4-3}. From a crystal-structure viewpoint, both TaIrTe$_4$ and NbIrTe$_4$ have a layered crystal structure, the two-dimensional (2D) atomic layer of which is composed of alternating quasi-1D IrTe$_2$ and (Ta,Nb)Te$_2$ chains. This structural feature is very common among ternary transition-metal tellurides, reminiscent of the ternary Pd-based tellurides Ta$_4$Pd$_3$Te$_{16}$ and Ta$_3$Pd$_3$Te$_{14}$, in both of which we reported superconductivity recently \cite{TaPdTe16,TaPdTe14}. The structural similarity and exotic properties
found among ternary transition-metal tellurides motivate us to further explore novel topological materials and even long-sought-after topological superconductors in these low-dimensional systems.

While a large number of materials have hitherto been identified to be topologically nontrivial, most of them are 2D or 3D electronically, and the quasi-1D analogues are extremely rare \cite{TaIrTe4-2,NbIrTe4-1,Bi4I4}. In quasi-1D materials, the competing ground states, such as charge/spin density waves or even superconductivity, often prevail and therefore elude the experimental detection of the topological carriers. Here, we report synthesis and characterizations of a new ternary Pd-based telluride TaPdTe$_5$, which explicitly shows a layered crystal structure with quasi-1D characteristics. Interestingly, the Shubnikov-de Haas (SdH) oscillations of the resistivity under high magnetic fields up to 51.7 T give the light effective masses of charge carriers. Our experiments also reveal a large unsaturated MR in the order of $\sim$ 10$^4$\% along with large anisotropy. From the analyses of de Haas-van Alphen (dHvA) oscillations, we have seen typical signatures of Dirac fermions in TaPdTe$_5$. The first-principles calculations further reveal topologically nontrivial states therein. Our results suggest the new material TaPdTe$_5$ hosts a topological nontrivial Berry phase and thus provides another platform to investigate topological physics in ternary transition-metal tellurides.

\section{Experimental methods}

\begin{table}
\caption{\label{LP}Crystallographic data and experiment details for TaPdTe$_5$.}
\begin{tabular}{l l l l}
\hline
\hline
    &  Compound	  & TaPdTe$_5$	  \\
\hline
&$T$ of data collection (K) &  296.02   &  \\
& Wavelength ({\AA}) &   0.71073 &  \\
& Formula weight (g mol$^{-1}$) &    925.37 &  \\
& Crystal system &  orthorhombic &  \\
& Space group &   $Cmcm$ &  \\
&$a$ ({\AA}) &  3.6934(4) & \\
&$b$ ({\AA}) & 13.2740(13) & \\
& $c$ ({\AA}) & 15.6020(15)  & \\
& Volume ({\AA}$^3$)&    764.91(13) &  \\
& Z &   4 &  \\
&Density (calculated) (g cm$^{-3}$) &    8.04&  \\
&$\mu$ (mm$^{-1}$)&  35.225   &  \\
&$F$(000) &    1516 &  \\
&Crystal Size (mm$^3$)&    0.23 $\times$ 0.12 $\times$ 0.05&  \\
& $\theta$ range for data collection (deg)&   2.611 to 26.475 \\
&Index ranges&  -4 $\leq$ $h$ $\leq$ 4 & \\
& &  -16 $\leq$ $k$ $\leq$ 16 & \\
& &  -19 $\leq$ $l$ $\leq$ 19 & \\
& Reflection collected&  3574 & \\
& Independent reflections &  479[$R_{int}$ = 0.0577] & \\
& Completeness to $\theta$ = 25.242$^\circ$ &  99.8\% & \\
& Data/restraints/parameters & 479/0/25   & \\
&Goodness of fit& 1.107   & \\
&Final $R^{\footnote{$R$ = $\Sigma ||F_o| - |F_c||/ \Sigma |F_o|$,
$\omega R = \{\Sigma[\omega(|F_o|^2-|F_c|^2)^2]/\Sigma[\omega(|F_o|^4)]\}^{1/2}$,
and $\omega$ = 1/$[\sigma^2(F_o^2)+(0.1162P)^2+(22.3597P)]$, where $P$ = ($F_o^2$ + 2$F_c^2$)/3.}}$ indices [$>$2$\sigma$($I$)]&  $R_{obs}$ = 0.0474 & \\
&&  $\omega R_{obs}$ = 0.1506 & \\
& $R$ indices (all data)&  $R_{all}$ = 0.0486  & \\
& &  $\omega R_{all}$ = 0.1526  & \\

\hline
\hline

\end{tabular}
\end{table}

\subsection{\label{subsec:level1}Sample synthesis}
Single crystals of TaPdTe$_5$ were grown using a self-flux method, similar to that in growing Ta$_4$Pd$_3$Te$_{16}$ and Ta$_3$Pd$_3$Te$_{14}$ single crystals \cite{TaPdTe16,TaPdTe14}, but the growth parameters were varied. The early attempt to synthesize TaPdTe$_5$ failed but made Ta$_3$Pd$_3$Te$_{14}$ accidently prepared \cite{TaNiTe5}. Powders of the elements Ta (99.97\%), Pd (99.995\%) and Te (99.99\%) used as reagents, in an atomic ratio of Ta : Pd : Te = 2 : 5 : 25, were thoroughly mixed together, loaded, and sealed into an evacuated quartz ampule. All the procedures handling the reagents were done in a glove box filled with highly pure argon gas. The ampule was slowly heated up to 1223 K and held for 24 h. After that, it was cooled to 923 K at a rate of 5 K/h, then followed by cooling down to room temperature. Large shiny gray-black flattened needle-like crystals with the dimensions up to 2 $\times$ 0.4 $\times$ 0.1 mm$^3$ were harvested (see the left inset of Fig.~\ref{fig1}(a)). The air-stable crystals are malleable, and can be easily exfoliated to a thin layer by a razor blade.

\subsection{\label{subsec:level2}Structure and Composition Determination}

\begin{figure*}
\begin{center}
\includegraphics[width=16cm]{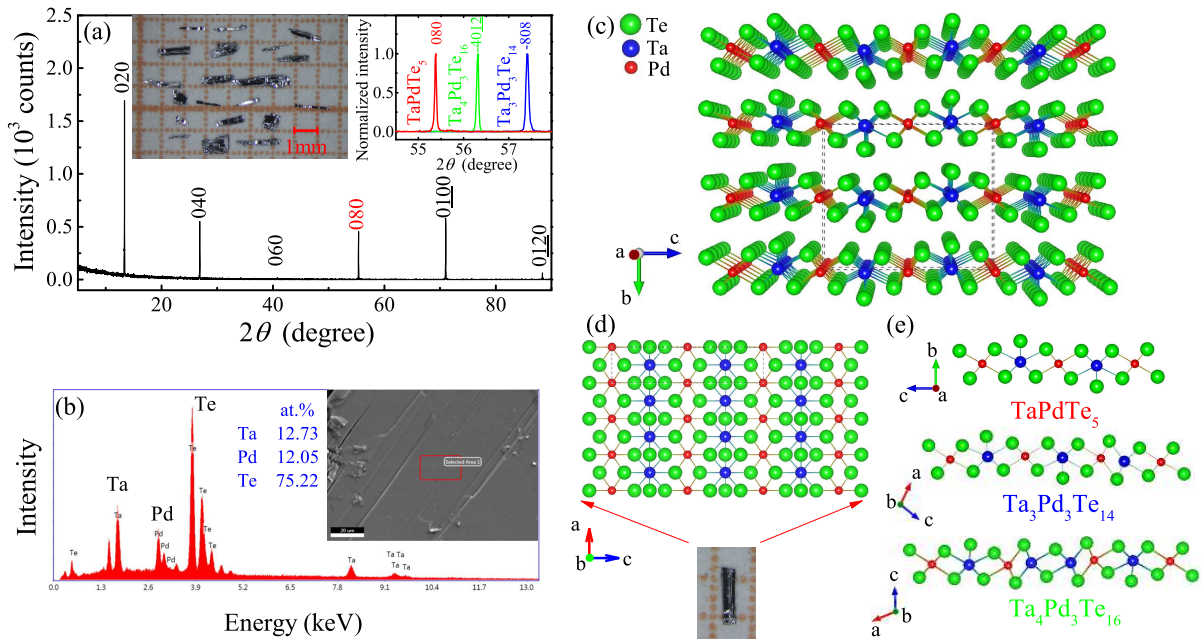}
\caption{\label{fig1}(Color online) Sample characterizations and crystallographic structure of TaPdTe$_5$. (a) Single-crystal X-ray diffraction pattern. The left inset is a
photograph of the TaPdTe$_5$ crystals on a millimeter-grid paper. The right inset shows the fourth reflection in X-ray diffraction pattern for TaPdTe$_5$, Ta$_4$Pd$_3$Te$_{16}$ and Ta$_3$Pd$_3$Te$_{14}$, respectively. (b) A typical energy-dispersive X-ray spectrum with electron beams focused on the selected area (marked in the inset) of the crystals. (c) Crystal structure of TaPdTe$_5$ viewed perspectively along the $a$-axis. (d) An individual atomic layer of TaPdTe$_5$ projected along the $b$-axis. The lower panel of (d) is a piece of the flattened needle-like crystal under an optical microscope, from which the layered and needle-like morphology can be clearly identified. (e) Projection view of one atomic layer of TaPdTe$_5$ (upper), Ta$_3$Pd$_3$Te$_{14}$ (middle) and Ta$_4$Pd$_3$Te$_{16}$ (bottom) along the chain direction.}
\end{center}
\end{figure*}

The chemical composition of the as-grown single crystals was measured by energy-dispersive X-ray spectroscopy (EDS) with an AMETEK\copyright EDAX (Model Octane Plus)
spectrometer, equipped in a field-emission scanning electron microscope (Hitachi S-4800). Figure~\ref{fig1}(b) shows the typical EDS spectrum for a piece of crystal. The average chemical composition of the as-grown crystals is determined to be Ta$_{1.00}$Pd$_{0.95}$Te$_{5.91}$. The obtained concentration of Te tends to be
larger than the actual one, possibly due to some systematic errors in the calculation process.

X-ray diffraction (XRD) data collection from this single crystal was performed on an Xcalibur, Atlas, Gemini ultra diffractometer. The crystal was mounted on the CCD goniometer head, followed by diffraction data collections at room temperature. Data collection, integration, and absorption correction were done in the X-area software package. The structure of TaPdTe$_5$ was solved by a direct method and refined by full-matrix least-squares based on $F^2$ using a SHELXTL program package \cite{Crystal}. The results indicate TaPdTe$_5$ adopts a $C$-centered orthorhombic structure with a space group $Cmcm$ (No. 63), which respects inversion symmetry. Atomic coordinates and thermal displacement parameters ($U_{eq}$) are given in Table S1 and Table S2 of the Supplemental Material (SM) \cite{SM}. The related bond lengths and bond angles are given in Table S3 of the SM. Single-crystal XRD data collection was also performed at room temperature with a monochromatic Cu K$_{\alpha 1}$ radiation using a PANalytical X-ray diffractometer (Model EMPYREAN) radiation by a conventional $\theta$-2$\theta$ scan for a crystal lying on a sample holder. As shown in Fig.~\ref{fig1}(a), only multiple peaks arising from the diffraction from (0 1 0) planes can be observed, consistent with the layered crystal structure of TaPdTe$_5$. The interplane spacing is determined to be 6.628 {\AA} by this XRD, and this value is very close to half of the lattice parameter $b$ ($b$/2 = 6.637 {\AA}) tabulated in Table~\ref{LP}. The layered nature of its structure is similar to that of the other two ternary Pd-based tellurides, Ta$_4$Pd$_3$Te$_{16}$ and Ta$_3$Pd$_3$Te$_{14}$ \cite{TaPdTe16,TaPdTe14}. To exhibit the obvious difference of the interlayer spacing among them, we plot the fourth reflection of XRD patterns together for TaPdTe$_5$, Ta$_4$Pd$_3$Te$_{16}$ and Ta$_3$Pd$_3$Te$_{14}$, namely (080), (-40\underline{12}) and (-808), respectively, in the right inset of Fig.~\ref{fig1}(a). By this means, one can easily distinguishes the TaPdTe$_5$ crystal from the other two, even though all these three crystals are of the same shiny gray-black strip-like exterior.

\subsection{\label{subsec:level3}Physical property measurements}

The resistivity and low-field magnetoresistivity measurements were carried out in a Physical Property Measurement System (PPMS-9, Quantum Design) with ac
transport option (typical excitation current 1 mA). A standard four-probe method was employed for the resistivity measurements. The Hall-effect measurement was performed by reversing the field direction and antisymmetrizing the data. Eight pieces of crystals with a total mass of 2.4 mg were used for the specific heat measurement on PPMS. High-field MR data up to 51.7 T were obtained from a pulsed high magnetic field equipment at Wuhan National High Magnetic Field Center. Magnetic susceptibility measurements were performed on a commercial Quantum Design magnetic property measurement system (MPMS-7).

\subsection{\label{subsec:level4}Band structure calculations}

We carried out first-principles calculations for band structure based on our experimental crystal structure. The electronic structure calculations with high accuracy were performed using the full-potential linearized augmented plane wave (FP$-$LAPW) method implemented in the WIEN2K code \cite{Wien2k}. The generalized gradient approximation (GGA) presented by Wu and Cohen \cite{GGA} was applied to the exchange-correlation potential calculation. The muffin tin radii were chosen to be 2.5 a.u.\ for all atoms. The plane-wave cutoff was defined by $RK_{max}=7.0$, where $R$ is the minimum LAPW sphere radius and $K_{max}$ is the plane-wave vector cutoff. To calculate the surface electronic
structure, we constructed a first-principles tight-binding model Hamilton, where the tight-binding model matrix
elements were calculated by projecting onto the Wannier orbitals \cite{wannier1}.
We used Ta $d$, Pd $d$ and Te $p$ orbitals to construct Wannier functions. The surface state spectrum of the (001) slab was calculated
with the surface Green's function methods as implemented in WannierTools \cite{wannier2}. Relativistic effects and SOC were included in the calculations.

\section{Results and Discussion}

The layered structure of TaPdTe$_5$ is displayed in Fig.~\ref{fig1}(c), which shows the eclipsed stacking of the layers along the $b$-axis. The layered slab, as shown in Fig.~\ref{fig1}(d), is composed of two alternating unique chains that run parallel to the $a$-axis. The chains are face-sharing Ta bicapped trigonal prisms and edge-sharing Pd octahedra. The types of coordination are common for the tellurides, such as two other ternary Pd-based tellurides Ta$_4$Pd$_3$Te$_{16}$ and Ta$_3$Pd$_3$Te$_{14}$. To compare the different constitution of the layered slab, we plot in Fig.~\ref{fig1}(e) the atomic layer of the three tellurides. It is interesting to note that we recently reported the superconductivity in both Ta$_4$Pd$_3$Te$_{16}$ and Ta$_3$Pd$_3$Te$_{14}$ \cite{TaPdTe16,TaPdTe14}. TaPdTe$_5$ is isostructural with NbNiTe$_5$ \cite{NbNiTe5} and Ta$TM$Te$_5$ ($TM$ = Ni, Pt) \cite{TaNiTe5,TaPtTe5}, for all of which the chains of transition metal atoms in one layer are aligned with the chains of transition metal atoms in adjacent layers in the $b$ direction. The same is true of the Ta or Nb chains. The other layered telluride NbPdTe$_5$, which has the same stoichiometry and almost identical atomic layer, is of a different structure type ($Pnma$), because of the different ordering of the layers being staggered rather than eclipsed \cite{NbPdTe5}. Due to the similar atomic radius, Nb and Ta usually have similar coordination preferences in tellurides. It is therefore rare that the substitution of Nb for Ta in NbPdTe$_5$ yields TaPdTe$_5$ with a despite similar but different structure.

\begin{figure}
\includegraphics[width=9cm]{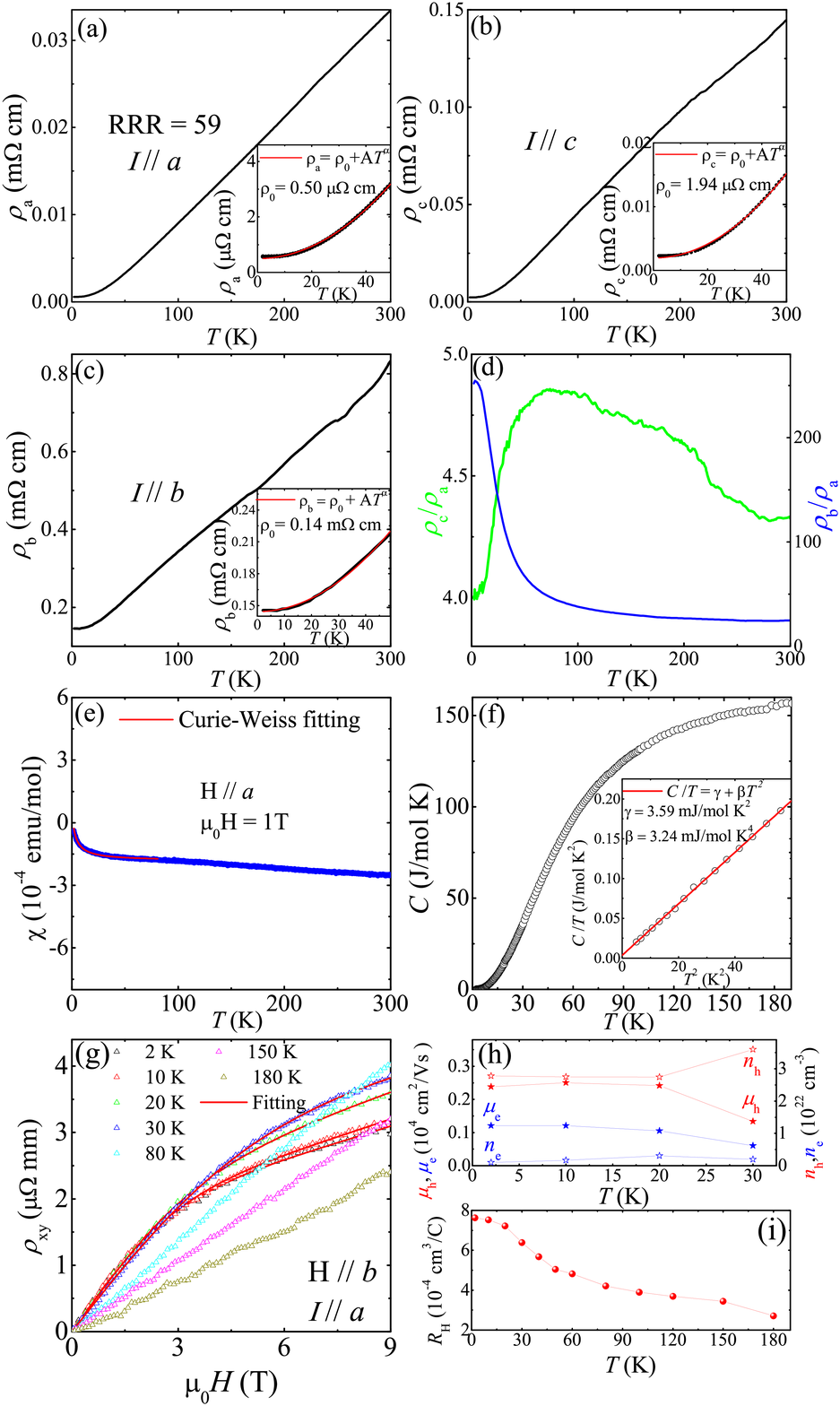}
\caption{\label{rt} (Color online) Physical properties of TaPdTe$_5$. (a, b, c) Temperature dependence of the electronic resistivity $\rho_a$, $\rho_c$ and $\rho_b$ with the current applied along the $a$-axis (a), $c$-axis (b) and $b$-axis (c), respectively. The insets of (a), (b) and (c) are an enlarged view of $\rho(T)$ in the low-temperature regime and the fit to $\rho(T)$ = $\rho_0$ + $A T^2$. (d) Temperature dependence of the resistivity anisotropy for $\rho_{c}$/$\rho_a$ and $\rho_{b}$/$\rho_a$. (e) Temperature dependence of the magnetic susceptibility $\chi (T)$, measured at the 1 T field parallel to the chain direction. (f) The heat capacity at zero field below 185 K. The inset separates the electronic and phononic contributions. (g) The Hall resistivity at different temperatures below 180 K. The red dashed lines are fits based on a two-band model. (h) Temperature dependence of the carrier density $n$ and the mobility $\mu$ extracted from the above fits. (i) The temperature dependence of low field Hall coefficients extracted by linear fitting.}
\end{figure}

The electrical resistivity, measured with the current applied along the $a$ [Fig.~\ref{rt}(a)], $c$ [Fig.~\ref{rt}(b)] and $b$ [Fig.~\ref{rt}(c)] directions, shows metallic behavior from room temperature to 2 K with a residual resistivity ratio RRR = $\rho_a$(300 K)/$\rho_a$(2 K) = 59. No indication of superconducting transition can be observed down to 0.5 K measured in a $^3$He cryostat (data not shown here). The room-temperature resistivity $\rho_a$(300 K) has a value of $\sim$ 33 $\mu \Omega$ cm, indicating a high degree of metallicity. From room temperature down to 50 K, $\rho_a$($T$), $\rho_c$($T$) and $\rho_b$($T$) follow an approximately linear temperature ($T$) dependence. Below 50 K, however, all of them cross over to
a quadratic $T$ dependence, $i.e.$, $\rho(T)$ = $\rho_0$ + $AT^2$, a Fermi-Liquid behavior, indicating the electron-electron scattering dominates in low-temperature region. The $T$ dependence of resistivity anisotropy $\rho_{c}$/$\rho_a$ and $\rho_{b}$/$\rho_a$ is plotted in Fig.~\ref{rt}(d). The quasi-1D transport behaviors are clearly exhibited with $\rho_a$ : $\rho_{b}$ : $\rho_{c}$ = 1 : 24.9 : 4.3 at 300 K, and 1 : 252.1 : 4 at 2 K. The small value of $\rho_{c}$/$\rho_a$ suggests a weak anisotropy in the layered slab, comparable with that of the superconductor Ta$_4$Pd$_3$Te$_{16}$ \cite{TaPdTe16-1}. The most striking feature of the data is that the sharp increase of the anisotropy for $\rho_{b}$/$\rho_a$ around 60 K, making the value at 2 K almost 1 order of magnitude larger than that at 300 K. A broad maximum also appears around 60 K for $\rho_{c}$/$\rho_a$. The origin of this anomaly is unclear at this stage. The magnetic susceptibility of TaPdTe$_5$  measured at a magnetic field of 1 T is shown in Fig.~\ref{rt}(c). It is apparent that $\chi(T)$ is almost $T$ independent in the high-temperature range. Below $T \sim$ 20 K, the magnetization undergoes a slight increase. $\chi (T)$ from 2 K to 80 K can be well described by the Curie-Weiss expression $\chi(T)$ = $\chi_0$ + $\frac{C}{T-\theta}$, where $\chi_0$ is the temperature-independent contribution, $C$ is the Curie constant and $\theta$ is the Curie-Weiss temperature. The fit gives $\chi_0$ = $-$ 1.82(1) $\times$ $10^{-4}$ emu/mol, $C$ = 8.17(4) $\times$ $10^{-4}$ emu K/mol and $\theta$ = 3.48(6) K. An effective moment calculated by the formula $\mu_{\textmd{eff}}$ = (8$C$)$^{1/2}$ gives 0.0809(1) $\mu_{\textmd{B}}$. Such a small $\mu_{\textmd{eff}}$, corresponding to $\sim$ 4.68 \% of that for an unpaired electron, indicates the Curie-Weiss contribution is most likely from tiny magnetic impurities. Therefore, TaPdTe$_5$ is expected to be paramagnetic, similar to TaPtTe$_5$ and NbPdTe$_5$ \cite{TaPtTe5}. By subtracting the Curie-Weiss-like contribution, we estimate the $T$ independent magnetic susceptibility as $\chi_0$ = $\chi_\textmd{p}$ + $\chi_{\textmd{vv}}$ + $\chi_{\textmd{L}}$ + $\chi_{\textmd{core}}$ = $-$ 1.82(1) $\times$ $10^{-4}$ emu/mol, which includes a Pauli paramagnetism ($\chi_\textmd{P}$), van Vleck paramagnetism ($\chi_{\textmd{vv}}$), Landau diamagnetic susceptibility ($\chi_{\textmd{L}}$) and core diamagnetism ($\chi_{\textmd{core}}$). The $\chi_{\textmd{L}}$ is estimated to be $\sim$ $-$ $\frac{1}{3}$$\chi_\textmd{p}$ by assuming $\chi_{\textmd{L}}$ being not enhanced by the electron-phonon interaction. $\chi_{\textmd{core}}$ is calculated to be $-$ 2.25(5) $\times$ $10^{-4}$ emu/mol from those of constituent ions in Ref. \cite{Core}. The upper limit of the Pauli paramagnetic susceptibility is then estimated to $\chi_\textmd{p}$ = 0.546 $\times$ $10^{-4}$ emu/mol by neglecting $\chi_{\textmd{vv}}$. The density of states at Fermi energy level $N(E_\textmd{F})$ could be derived to be 1.69 eV$^{-1}$ fu$^{-1}$ from the formula $N(E_\textmd{F})$ = $\frac{\chi_\textmd{p}}{\mu_0\mu_{\mathrm{B}}}$, where $\mu_0$ and $\mu_{\mathrm{B}}$ denote the vacuum permeability and Bohr magneton, respectively. We can also obtain the Sommerfield parameter $\gamma_n$ = 3.99 mJ mol$^{-1}$ K$^{-2}$ by the relation
$\gamma_n$ = $k_{\textmd{B}}^2 \pi^2 N(E_\textmd{F})$/3 for noninteracting electron systems, where $k_{\textmd{B}}$ is the Boltzmann constant.

\begin{figure*}
\includegraphics[width=16cm]{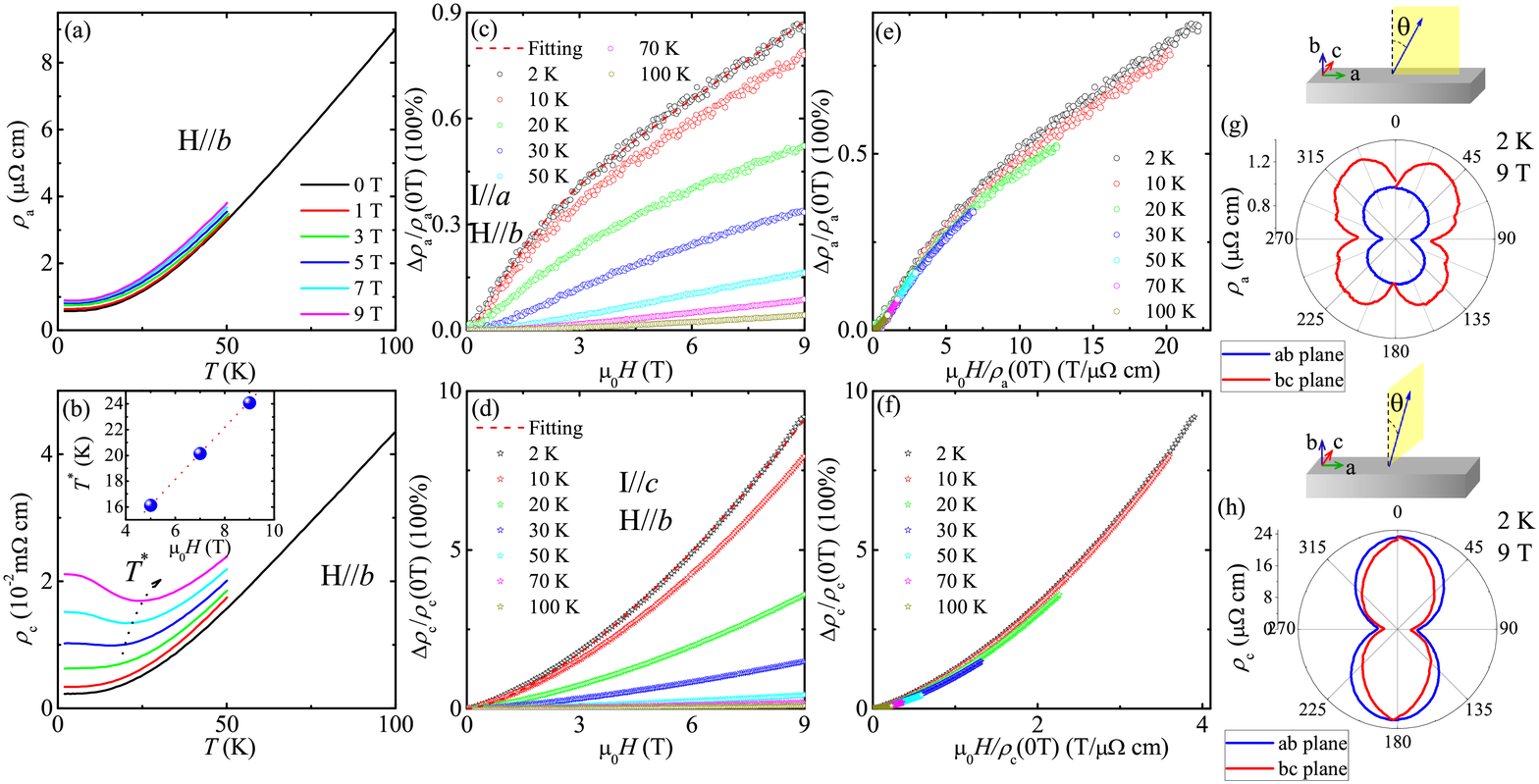}
\caption{\label{mr} (Color online)  Temperature dependence of $\rho_a$ (a) and $\rho_c$ (b) with the magnetic fields parallel to the $b$-axis up to 9 T below 50 K. The inset of (b)
is the field dependence of $T^*$. The MR for $\rho_a$ (c) and $\rho_c$ (d) at selected temperatures. The red dashed lines are the fits (see main text). Kohler's scaling for $\rho_a$ (e) and $\rho_c$ (f). Angular dependence of MR at 2 K for $\rho_a$ (g) and $\rho_c$ (h) with magnetic field 9 T being rotated within the $ab$ (blue) and $bc$ (red) planes. The inset shows the definition of field orientation angle $\theta$.}
\end{figure*}

The $T$-dependent specific heat, as shown in Fig.~\ref{rt}(d), shows no evident anomaly below 190 K. The data below 8 K can be well described by the equation $C$ = $\gamma T$ + $\beta T^3$ shown in the inset of Fig.~\ref{rt}(d), where $\gamma T$ represents the electron contribution and $\beta T^3$ represents the phonon contribution. The fit yields $\gamma$ = 3.59 mJ mol$^{-1}$ K$^{-2}$, close to the upper limit of $\gamma_n$ as analysed above, and $\beta$ = 3.24 mJ mol$^{-1}$ K$^{-4}$, from which the Debye temperature $\Theta_\mathrm{D}$ is estimated to be 161.39 K. The Wilson ration $R_\textmd{W}$ [= $\pi^2k_{\mathrm{B}}^2\chi_\textmd{P}$/3$\mu_0\mu_{\mathrm{B}}^2\gamma$] can measure the relative enhancements of the spin susceptibility and electronic specific heat \cite{Wilson}. To estimate the strength of the electron-electron correlation effect, we calculate the Wilson ratio is $\sim$ 1.1, very close to the value ($R_\textmd{W}$ = 1) of non-interacting Fermi liquid. This suggests that the electron-electron correlation effect is negligible in TaPdTe$_5$.

The Hall resistivity $\rho_{xy}$ up to 9 T at selected temperatures below 180 K is shown in Fig.~\ref{rt}(e), from which nonlinear field-dependent $\rho_{xy}$ can be clearly observed, reflecting the multiband characteristic in TaPdTe$_5$. The positive value of $\rho_{xy}$ indicates the dominance of hole-type charge carriers. We apply a simplified two-carrier model including one electron-type band and one hole-type band, in an effort to extract the charge carrier densities and mobilities. According to the classical expression for the Hall coefficient, including both electron- and hole-type carriers \cite{Classical-hall}:

\begin{align}
\frac{\rho_{xy}}{\mu_0H} =\frac{1}{e}\frac{(\mu_h^2n_h-\mu_e^2n_e)+(\mu_h\mu_e)^2(\mu_0H)^2(n_h-n_e)}{(\mu_en_h+\mu_hn_e)^2+(\mu_h\mu_e)^2(\mu_0H)^2(n_h-n_e)^2},
\end{align}
where $e$ is the absolute value of electronic charge, $n_{e(h)}$ and $\mu_{e(h)}$ are the carrier densities and mobilities of the electrons (holes), respectively. This equation predicts
immediately  a nonlinear field dependence, once two carrier types are present. The fit is well performed by the nonlinear data below 30 K. It is apparent that the hole density $n_{h}$ (2.78 $\times$ 10$^{22}$ cm$^{-3}$ at 2 K) is much higher than the electron density $n_{e}$ (9.43 $\times$ 10$^{20}$ cm$^{-3}$ at 2 K). Moreover, the holes have much higher mobility (2.38 $\times$ 10$^{3}$ cm$^{2}$ V$^{-1}$ s$^{-1}$ at 2 K) than the electrons (1.21 $\times$ 10$^{3}$ cm$^{2}$ V$^{-1}$ s$^{-1}$ at 2 K). Both $n_{e(h)}$ and $\mu_{e(h)}$ vary not too much within this temperature interval. At high temperatures ($T$ $>$ 30 K), $\rho_{xy}$ is still positive and develops linearly with the field, indicating the single hole-type carriers dominate the transport. Clearly, this result excludes the possibility of electron-hole compensation effect in TaPdTe$_5$. The $T$-dependent Hall coefficient $R_{\textmd{H}}$ is plotted in Fig.~\ref{rt}(f), obtained by linear fitting with the $\rho_{xy}$ below the field 1.5 T. The obvious $T$ dependent $R_{\textmd{H}}$($T$) is also
signaling multiple bands with coexisting hole and electron pockets, consistent with the above analysis, as each band can have a distinct $T$-dependent mobility.

The MR of $\rho_a$ under magnetic field $\emph{\textbf{H}}$ parallel to the $b$ direction up to 9 T is shown in Fig.~\ref{mr}(a), from which we can see that the $T$ evolutions of $\rho_a$ under different fields are almost identical. We also measured the field dependent MR of $\rho_a$ at several fixed temperatures, defined as $\Delta \rho_a$/$\rho_a$(0 T) [$\Delta \rho_a$ = $\rho_a$ $-$ $\rho_a$(0 T)]. As noticed in Fig.~\ref{mr}(c), the MR of $\rho_a$ is weak, reaching only 87\% at 2 K and 9 T. Importantly, its field dependence is quite linear at low temperatures, a feature seen in many topological materials due to the distinct spectrum of Landau levels for Dirac fermions in the field \cite{Cd3As2,PdSn4}, while a seemingly quadratic contribution can be observed in low field. For a standard metal with the presence of both electrons and holes, a quadratic $T$ dependence is expected at low field and the MR tends to saturate at high field, $i. e.$, MR should scale as $\frac{\alpha(\mu_0H)^2}{\beta + (\mu_0H)^2}$ \cite{Metal}. Hence, the overall MR data with both Dirac fermions and conventional carriers can be fitted by a linear term in addition to a conventional component, $i.e.$, $\frac{\Delta \rho_a}{\rho_a}$ = $\frac{\alpha(\mu_0H)^2}{\beta + (\mu_0H)^2}$ + $\gamma \mu_0H$. As shown in Fig.~\ref{mr}(c), the experimental data at 2 K can be fitted very well. The MR of $\rho_c$ under $\emph{\textbf{H}}\parallel b$ is rather large, reaching as high as 910\% at 2 K and 9 T, nearly 11 times that of $\rho_a$, as clearly illustrated in Fig.~\ref{mr}(c). The MR of $\rho_c$ can be fitted to a single power law  $\Delta \rho_c$($\mu_0 H$) $\propto$ $H^\alpha$ with $\alpha$ = 1.50. More interestingly, with increasing the field above 5 T, the $T$ evolution of $\rho_c$ no longer follows the zero-field curve, as shown in Fig.~\ref{mr}(b). Below a 'turn on' temperature $T^*$, defined as the temperature at which the minimum in the resistivity is located, $\rho_c$ begins to increase and finally saturate to a plateau at low temperatures. Very similar resistivity plateaus have been observed in a number of topological semimetals, and have been attributed by some groups to the presence of the topological surface states after excluding the possibility of a magnetic-field-driven metal-insulator transition \cite{YSb,Li}. The 'turn on' temperature is linearly shifted to higher temperature as larger fields are applied at the rate of $\sim$ 2.0 K T$^{-1}$, as shown in the inset of Fig.~\ref{mr}(b), suggesting competition between dominating scattering mechanisms. The linear dependence of $T^*$ on magnetic fields is also reported in topological material WTe$_2$ \cite{Ali}. On the other hand, the MR of many metals and semimetals obeys a general function, commonly referred to as Kohler's rule, $\Delta \rho$/$\rho_0$ = $f(H/\rho_0)$,
where $\rho_0$ is the zero-field resistivity. As shown in Fig.~\ref{mr}(e) and Fig.~\ref{mr}(f), the MR curves of both $\rho_a$ and $\rho_c$ can basically be scaled into a single
curve, indicating that Kohler's rule is well obeyed over a large temperature range. The validation of Kohler's rule indicates that the upturns and low-temperature plateau of the MR  do not reflect an intrinsic field-induced metal-insulator transition, but are rather a consequence of the small residual resistivity and correspondingly high mobilities of the charge carriers \cite{Turn-on}.

\begin{figure}
\includegraphics[width=7cm]{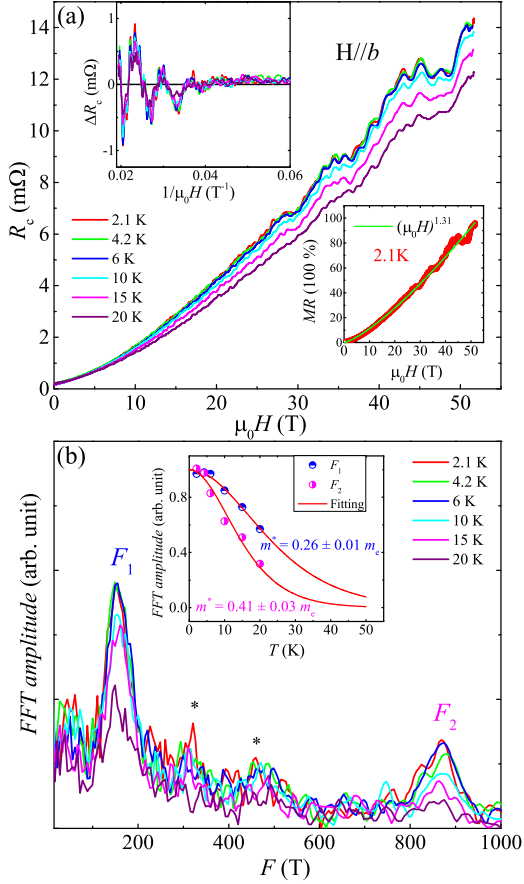}
\caption{\label{hf} (Color online) (a) The high field MR for $\rho_c$ under $\emph{\textbf{H}}\parallel b$ at different temperatures starting from 2.1 K to 20 K. The upper inset shows SdH quantum oscillation after background subtraction. The lower inset shows the normalized MR at 2.1 K. (b) Corresponding FFT amplitudes of SdH oscillations depicting fundamental
frequencies $F_1$ (150 T) and $F_2$ (870 T). Two small peaks are marked by the asterisk. The inset shows $T$-dependent FFT amplitudes of $F_1$ (150 T) and $F_2$ (870 T) along with their fittings according to Lifshitz-Kosevich relation to obtain their corresponding effective masses.}
\end{figure}

The anisotropic behavior in the MR can be visualized in the angular dependence of magnetoresistivity (AMR). The AMR of $\rho_a$ at 2 K and 9 T is plotted in Fig.~\ref{mr}(g). For fields rotating within the $bc$ plane, $\rho_a$($\theta$) reaches its maximum for $\theta$ $\approx$ 45$^\circ$ and its minimum for $\theta$ = 0$^\circ$ and 90$^\circ$, resulting in a fourfold symmetric "butterfly" shaped angular dependence. This butterfly shaped MR was also observed in a few of materials including high-$T_c$ superconductor and topological semimetals, for example, type-II Weyl semimetal NbIrTe$_4$ \cite{NbIrTe4-1} and Dirac nodal line semimetal ZrSiS \cite{ZrSiS}. The unconventional fourfold anisotropy is believed to be governed by the topography of Fermi surface and related anisotropy in effective masses in NbIrTe$_4$ \cite{NbIrTe4-1}. In the case of ZrSiS, this behavior is ascribed to a topological phase transition as a function of field orientation that is inherent to the nodal Dirac line \cite{ZrSiS}. For fields rotating within the $ab$
plane, the AMR data of $\rho_a$ exhibit a typical twofold symmetric "dumbbell" shaped anisotropy, which is expected for a material with 2D or quasi-2D electronic structure and classical Lorentz-type MR, $i.e.$ AMR $\propto$ ($H$cos$\theta$)$^2$. The possible presence of 2D-like Fermi surface is in line with the layered crystal structure of TaPdTe$_5$.
The AMR of $\rho_c$ at 2 K and 9 T, for fields rotating either within the $ab$ plane or within the $bc$ plane, also shows a "dumbbell" shape with twofold symmetry, as noticed in Fig.~\ref{mr}(h). The AMR at a fixed angle within the $ab$ plane, is a little larger than that within the $bc$ plane, resulting in a fatter dumbbell.

\begin{figure*}
\includegraphics[width=15cm]{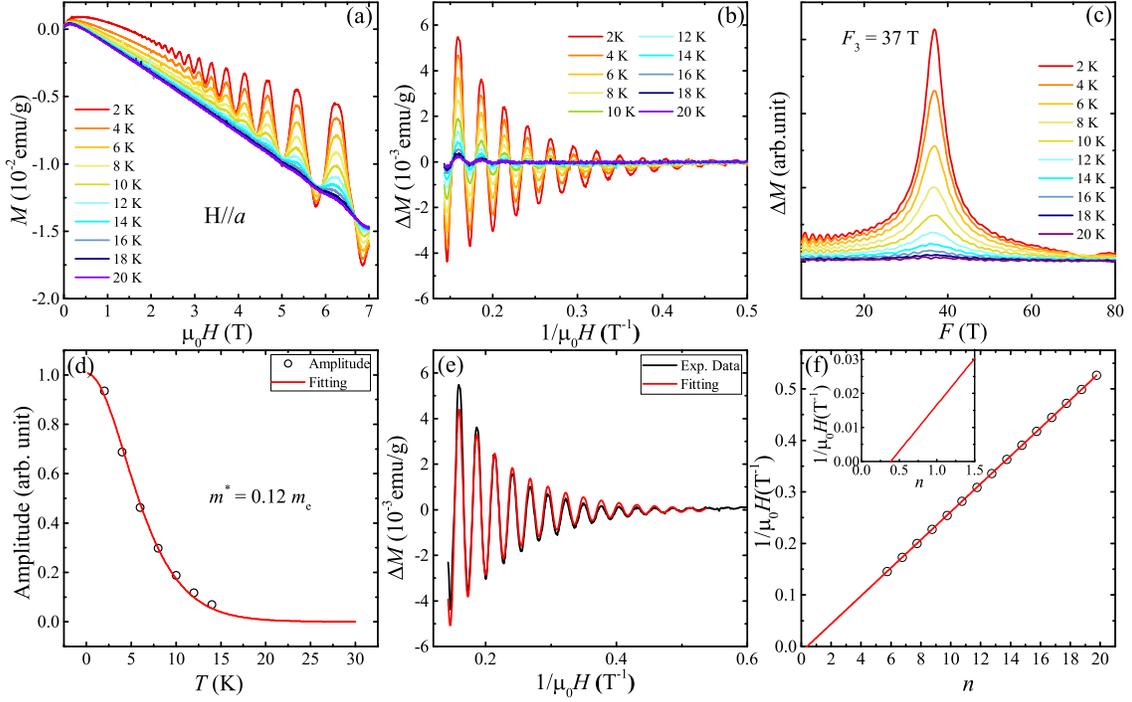}
\caption{\label{dHvA} (Color online) The dHvA oscillations and nontrivial Berry phase. (a) Isothermal magnetization under $\emph{\textbf{H}}\parallel a$ at different temperatures starting from 2 K to 20 K. (b) The magnetization oscillations at different temperatures after subtracting the background. (c) The corresponding FFT spectrum. (d) The FFT amplitude as a function of temperature and the fit to $R_\textmd{T}$ to determine the effective mass. (e) The Lifshitz-Kosevich fit (red line) of the oscillation pattern (black line) at 2 K. (f) The Landau's fan diagram for the identified frequency. The inset enlarges the intersection.}
\end{figure*}

To check whether there is any SdH quantum oscillation in TaPdTe$_5$, we measured the MR under the high magnetic field $\emph{\textbf{H}}\parallel b$ up to 51.7 T. The resistivity signal measured with the current along the $a$-axis is indistinguishable. It seems that the resistivity signal is submerged by noisy signal, possibly due to the resistivity $R_a$ being quite small. The MR with the current along the $c$-axis is shown in Fig.~\ref{hf}(a), from which SdH oscillations can be identified above 28.5 T. To extract their amplitude of the oscillations, a third-order polynomial background is subtracted from the resistivity. The resulting oscillations at different temperatures are plotted against the inverse magnetic field, as seen in the upper inset of Fig.~\ref{hf}(a). Applying a fast Fourier transform (FFT), we obtained the corresponding frequency spectra, as plotted in Fig.~\ref{hf}(b). Although the data is not smooth enough, two major frequencies 150 T and 870 T, which are denoted as $F_1$ and $F_2$, respectively, can be clearly identified. The observed frequencies are related to the extremal cross sections of the
Fermi surface ($A_\textmd{F}$) described by the Onsager relation, $F_i$ = $A_i\hbar$/(2$\pi e$), where $\hbar$ is the reduced Planck constant. The
calculated $A_i$ are $A_1$ = 1.43 nm$^{-2}$ and $A_2$ = 8.29 nm$^{-2}$. The $T$ dependence of the FFT amplitudes of two corresponding frequencies is shown in
the inset of Fig.~\ref{hf}(b). The cyclotron effective mass of a given electronic orbit can be extracted from the temperature damping factor $R_T$ in the Lifshitz-Kosevitch (LK) formula and is given by $R_\textmd{T}$ = $\frac{\lambda T}{sinh(\lambda T)}$ where $\lambda$ = 2$\pi^2k_\textmd{B}m^*$/$\hbar e \mu_0H$, with $m^*$ being the cyclotron effective mass and $k_\textmd{B}$ being the Boltzmann constant. The resultant $m^*$ are (0.26 $\pm$ 0.01)$m_e$ and (0.41 $\pm$ 0.03)$m_e$ for $F_1$ and $F_2$, respectively, where $m_e$ is the free electron mass. The lower inset of Fig.~\ref{hf}(a) shows the MR of $R_c$. The high field MR at 2.1 K can be fitted to a power law where $\Delta R_c$ $\propto$ $H^\alpha$ with
$\alpha$ = 1.31, close to the derived value 1.50 from fitting the low field MR. This indicates the field dependence of MR basically holds in high field range. Moreover, the MR does
not show any sign of saturation and reaches a magnitude of (9.5 $\times$ 10$^3$)\% at 51.7 T. In a classical picture, the large nonsaturating transverse MR is due to the vicinity to
a perfect balance between electron and hole carriers \cite{Ali}. Our analysis on the data from Hall-effect measurement excludes the electron-hole compensation explanation for the
nonsaturating MR, making us conjecture that it might bear relation to the existence of unconventional quasiparticles in TaPdTe$_5$.

\begin{figure*}
\includegraphics[width=12cm]{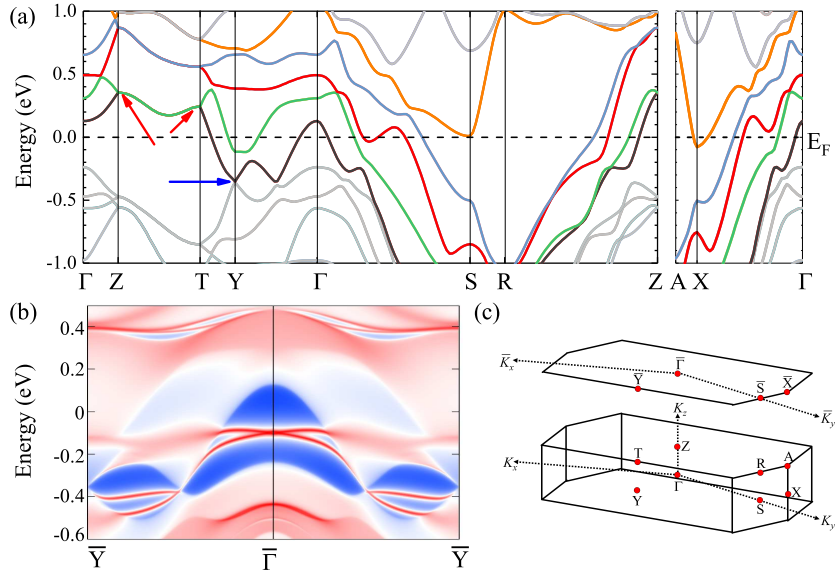}
\caption{\label{band} (Color online) (a) Calculated band structure for TaPdTe$_5$ with spin-orbit coupling included. The bands crossing the Fermi level are marked by different colors. The red and blue arrows indicate the positions of Dirac points. (b) Surface state spectrum of TaPdTe$_5$. (c) The bulk Brillouin zone (BZ) and (001) surface BZ of TaPdTe$_5$ with high-symmetry points marked in red. }
\end{figure*}

While SdH oscillations can only be observed in the high field MR when $\mu_0 H$ exceeds 28.5 T, dHvA effect, the quantum oscillations in magnetic susceptibility, to be discussed below, appear in a magnetic field as low as 1.7 T at 2 K. This inconsistence of SdH and dHvA oscillations is often observed in low-dimensional materials due to the distinct mechanisms of SdH and dHvA oscillations \cite{NiTe2}.
The isothermal magnetization for $\emph{\textbf{H}}\parallel a$ at several temperatures below 20 K shows beautiful quantum oscillations [Fig.~\ref{dHvA}(a)].
We present the oscillatory components of magnetization, obtained after subtracting the background, in Fig.~\ref{dHvA}(b). Very strong oscillations are clearly seen.
From the FFT analyses, we derive a single frequency of 37 T from the dHvA oscillations for $\emph{\textbf{H}}\parallel a$, as shown in Fig.~\ref{dHvA}(c).
The Fermi area ($A_F$) that is covered by electrons or holes is found to be 0.352 nm$^{-2}$ from the Onsager relation. In general, the
oscillatory dHvA data can be described by the standard LK formula \cite{LK}, with the Berry phase being taken into account \cite{Berry}:
\begin{align}
\Delta M \propto - R_T \cdot R_D \cdot R_S \cdot sin[2\pi(\frac{F}{\mu_0H} + \frac{1}{2} - \frac{\varphi_B}{2\pi} - \delta)],
\end{align}
where $\varphi_\textmd{B}$ is the Berry phase, $\delta$ is an additional phase shift determined by the dimensionality of the Fermi surface, $R_\textmd{T}$, $R_\textmd{D}$, and
$R_\textmd{S}$ are the thermal damping factor, Dingle damping term, and a spin-related damping term, respectively.
$R_\textmd{T}$ = $\frac{2\pi^2k_\textmd{B}m^* T/\hbar e \mu_0H}{sinh(2\pi^2k_\textmd{B}m^* T/\hbar e \mu_0H)}$, $R_\textmd{D}$ = exp(2$\pi^2k_\textmd{B}T_\textmd{D}m^*$/$\hbar e\mu_0H$), and $R_\textmd{S}$ = cos($\pi gm^*$/2$m_e$), where $T_\textmd{D}$ is the Dingle temperature. The oscillation of $\Delta M$ is described by the sine term with the phase factor $\frac{1}{2} - \frac{\varphi_B}{2\pi} - \delta$. The phase shift $\delta$ is equal to 0 and $\pm$ 1/8 ($-$ for electronlike and $+$ for the holelike), respectively, for the 2D and 3D Fermi surfaces.
As we have done above, the effective mass $m^*$ can be obtained through the fit of the $T$ dependence of the oscillation amplitude to the thermal damping
factor $R_\textmd{T}$ from the LK formula. The effective mass is extracted to be 0.12$m_e$, as shown in Fig.~\ref{dHvA}(d). Using the fitted effective mass as a known parameter, we can further fit the oscillation pattern at 2 K, represented by the black line in Fig.~\ref{dHvA}(e), to the LK formula, from which quantum mobility and the Berry phase can be extracted.
It is well known that the Berry phase is zero for a parabolic energy dispersion and $\pi$ for a linear energy dispersion.
Given that the Fermi surface of TaPdTe$_5$ is of 2D and/or 3D character, we adopt both all possible values of $\delta$. The fit gives the Berry phase is 1.24$\pi$, 1.49$\pi$ and
0.99$\pi$ for $\delta$ = 0, $-$ 1/8 and 1/8, respectively. The fitted Ding temperature $T_\textmd{D}$ is 5.85 K, corresponding to the quantum relaxation time $\tau_q$ = $\hbar$/(2$\pi k_\textmd{B}T_\textmd{D}$) = 2.08 $\times$ 10$^{-13}$ s and quantum mobility $\mu_q$ = $e\tau$/$m^*$ = 3159 cm$^2$ V$^{-1}$ s$^{-1}$.
The Berry phase can also be extracted from the Landau's fan diagram as shown in Fig.~\ref{dHvA}(f) \cite{ZrSiSe}. The valley in the magnetization should be assigned
with a Landau level (LL) index of $n$ $-$ 1/4. The red line in Fig.~\ref{dHvA}(f) is the linear fit of the LL indices. The slope of the linear fit is 36.8 T, in good agreement with the frequency obtained from the FFT analysis. The inset of Fig.~\ref{dHvA}(f) is an enlarged view near 1/$\mu_0H$$\rightarrow$ 0. The intercept at the $n$ abscissa from the linear
fit is 0.39 $\pm$ 0.03, corresponding to a nontrivial Berry phase of $\varphi_\textmd{B}$ = 2$\pi|0.39 + \delta|$. A nearly 1/2-shift possibly distinguishes the Dirac spectrum from the Schr\"{o}dinger case for the electronic states in TaPdTe$_5$ \cite{Shift}. Thus, the fitting parameters obtained from different methods
consistently verify the nontrivial Berry phase in TaPdTe$_5$.

To further identify the topological electronic properties, we performed the full-potential linearized augmented plane wave method implemented in the WIEN2k package \cite{Wien2k} for the electronic structure calculations, taking account of the SOC. The GGA presented by Wu and Cohen \cite{GGA} was used for the exchange-correlation energy calculations. The calculated band structure is shown in Fig. \ref{band}(a), with five bands crossing the Fermi level. Similar to the quasi-one-dimensional Tl$_2$Mo$_6$Se$_6$ compound \cite{Tl2Mo6Se6,Cubic}, TaPdTe$_5$ also possesses cubic Dirac crossings, which have linear band crossings along one principle axis but the cubic dispersions in the plane perpendicular to it. The red arrows indicate the cubic Dirac crossings at high symmetry points: Z (0, 0, 0.5) and T ($-$0.5, 0.5, 0.5), and the blue arrow indicates the standard linear Dirac corssing at Y ($-$0.5, 0.5, 0). $\mathbb{Z}_2$ topological indices ($\nu_0$; $\nu_1$$\nu_2$$\nu_3$) are usually used in the classification of topological band insulators and semimetals \cite{Fu2007}. Therefore, we have calculated the $\mathbb{Z}_2$ topological number, which is valid for time-reversal invariant system. A tight-binding model based on Wannier functions \cite{wannier1,wannier2} was constructed to obtain the topological properties, using Ta $d$, Pd $d$, and Te $p$ orbitals with SOC included. The $\mathbb{Z}_2$ topological number for 3D bulk system can be obtained from the calculation of the Wilson loop for the six time-reversal invariant momentum plane. The $\mathbb{Z}_2$ invariants of TaPdTe$_5$ are 1 for $k_x$ = $\pi/a$, $k_y$ = $\pi/b$, $k_z$ = 0 plane, while zeros for other planes. The topological index is (1; 110), which indicates that TaPdTe$_5$ is a topological nontrivial material. From the surface state spectrum in Fig. \ref{band}(b), there is a surface Dirac cone at the $\bar{\Gamma}$ point, which is also characteristic of a strong topological material \cite{zhang2009}. More detailed results and discussions of first-principle calculations will be given elsewhere in a separated paper.

\section{Summary}
In summary, we have successfully synthesized the high quality single crystal of a new layered ternary telluride TaPdTe$_5$ with quasi-1D PdTe$_2$ chains. Hall resistivity measurement indicates the coexistence of highly mobile, but unbalanced electron and hole carriers. Anisotropic MR behaviors are found for magnetic fields and currents applied along different crystallographic axes. The MR for $\emph{\textbf{I}}\parallel a$ and $\emph{\textbf{H}}\parallel b$ can be separated
into a linear term in addition to a conventional component, expected for Dirac fermions and conventional carriers, respectively, while the MR for $\emph{\textbf{I}}\parallel c$ and $\emph{\textbf{H}}\parallel b$ up to 51.7 T exhibits a power-law dependence even in high field range. The anomalous nonsaturating high MR of $\sim$ 10$^4$\% at 2.1 K and 51.7 T for $\emph{\textbf{I}}\parallel c$ possibly bears the relation to unconventional quasiparticles in TaPdTe$_5$. SdH oscillations under high magnetic fields give the light effective masses of charge carriers. More interestingly, clear dHvA oscillationsreveal a nontrivial Berry phase in TaPdTe$_5$. The first-principle calculations indeed verify TaPdTe$_5$ is a topological nontrivial material and possesses ordinary Dirac crossing as well as cubic Dirac crossing. As we have demonstrated in the initial trial, its atomically thin layers can be easily obtained through mechanical exfoliation. Therefore, the new layered telluride TaPdTe$_5$ provides a new platform to study the nontrivial physics belonging to "relativistic" carriers in low dimensions. Angle dependent dHvA oscillations and angle-resolved photoemission spectroscopy are urgently called for to provide further insights into its Fermiology and the topological nature and identify the contributing pocket of each frequency. In the future, it would be also intriguing to explore the possible topological superconductivity in TaPdTe$_5$ by chemical substitution, intercalation, or the high pressure.

\begin{center}
{\bf ACKNOWLEDGEMENTS}
\end{center}

The authors would like to thank J. K. Bao for his constructive suggestions on the single-crystal diffraction experiment. Thanks are also due to S. Y. Li for
fruitful discussions. This work was supported by Zhejiang Provincial Natural Science Foundation of China (No. LY19A040002). X. F. Xu would like to acknowledge the financial support from the National Natural Science Foundation of China (Nos. 11974061 and U1732162). B. Li was supported by the National Natural Science Foundation of China (No. 11674054) and NUPTSF (Nos. NY220038 and NY219087). G. H. Cao acknowledges the supports from the National Key Research and Development Program of China (2017YFA0303002 and 2016YFA0300202).



\begin{thebibliography}{00}
\bibitem{Kane} M. Z. Hasan and C. L. Kane, Colloquium: Topological insulators, Rev. Mod. Phys. \textbf{82}, 3045 (2010).
\bibitem{Zhang} X. L. Qi and S. C. Zhang, Topological insulators and superconductors, Rev. Mod. Phys. \textbf{83}, 1057 (2011).
\bibitem{Wan} X. Wan, A. M. Turner, A. Vishwanath, and S. Y. Savrasov, Topological semimetal and Fermi-arc surface states in the electronic structure of pyrochlore iridates, Phys. Rev. B 83, 205101 (2011).
\bibitem{Wang} Z. Wang, Y. Sun, X. Qiu Chen, C. Franchini, G. Xu, H. Weng, X. Dai, and Z. Fang, Dirac semimetal and topological
phase transitions in A$_3$Bi (A = Na, K, Rb), Phys. Rev. B \textbf{85}, 195320 (2012).
\bibitem{Chen1} Z. K. Liu, B. Zhou, Y. Zhang, Z. J. Wang, H. M. Weng, D. Prabhakaran, S. K. Mo, Z. X. Shen, Z. Fang, X. Dai, Z. Hussain, and Y. L. Chen, Discovery of a Three-Dimensional Topological Dirac Semimetal Na$_3$Bi, Science \textbf{343}, 864 (2014).
\bibitem{Chen2} Z. K. Liu, J. Jiang, B. Zhou, Z. J. Wang, Y. Zhang, H. M. Weng, D. Prabhakaran, S. K. Mo, H. Peng, P. Dudin, T. Kim, M. Hoesch, Z. Fang, X. Dai, Z. X. Shen, D. L. Feng, Z. Hussain, and Y. L. Chen, A stable three-dimensional topological Dirac semimetal Cd$_3$As$_2$, Nat. Mater. \textbf{13}, 677 (2014).
\bibitem{Dai} H. Weng, C. Fang, Z. Fang, B. A. Bernevig, and X. Dai, Weyl Semimetal Phase in Noncentrosymmetric Transition-Metal Monophosphides, Phys. Rev. X \textbf{5}, 011029 (2015).
\bibitem{Chen3} Z. K. Liu, L. X. Yang, Y. Sun, T. Zhang, H. Peng, H. F. Yang, C. Chen, Y. Zhang, Y. F. Guo, D. Prabhakaran, M. Schmidt, Z. Hussain, S. K. Mo, C. Felser, B. Yan, and Y. L. Chen, Evolution of the Fermi surface of Weyl semimetals in the transition metal pnictide family, Nat. Mater. \textbf{15}, 27 (2016).
\bibitem{Xu} S. Y. Xu, I. Belopolski, N. Alidoust et al., Discovery of a Weyl Fermion semimetal and topological Fermi arcs,
Science \textbf{349}, 613 (2015).
\bibitem{Huang} S. M. Huang, S. Y. Xu, I. Belopolski, C. C. Lee, G. Chang, B. Wang, N. Alidoust, G. Bian, M. Neupane, C. Zhang, S. Jia, A. Bansil, H. Lin, and M. Z. Hasan, A Weyl Fermion semimetal with surface Fermi arcs in the transition metal monopnictide TaAs class, Nat. Commun. \textbf{6}, 7373 (2015).
\bibitem{Noh} H. J. Noh, J. Jeong, E. J. Cho, K. Kim, B. I. Min, and B. G. Park, Experimental Realization of Type-II Dirac Fermions in a PdTe$_2$ Superconductor, Phys. Rev. Lett. \textbf{119}, 016401 (2017).
\bibitem{Fei} F. C. Fei, X. Y. Bo, R. Wang, B. Wu, J. Jiang, D. Z. Fu, M. Gao, H. Zheng, Y. L. Chen, X. F. Wang, H. J. Bu, F. Q. Song, X. G. Wan, B. G. Wang, and G. H. Wang, Nontrivial Berry phase and type-II Dirac transport in the layered material PdTe$_2$, Phys. Rev. B \textbf{96}, 041201(R) (2017).
\bibitem{Bernevig} A. A. Soluyanov, D. Gresch, Z. Wang, Q. Wu, M. Troyer, X. Dai, and B. A. Bernevig, Type-II Weyl semimetals, Nature (London) \textbf{527}, 495 (2015).
\bibitem{Bian} G. Bian, T. R. Chang, R. Sankar et al., Topological nodal-line fermions in spin-orbit metal PbTaSe$_2$, Nat. Commun. \textbf{7}, 10556 (2016).
\bibitem{Mao} J. Hu, Z. J. Tang, J. Y. Liu, X. Liu, Y. L. Zhu, D. Graf, K. Myhro, S. Tran, C. N. Lau, J. Wei, and Z. Q. Mao, Evidence of Topological Nodal-Line Fermions in ZrSiSe and ZrSiTe, Phys. Rev. Lett. \textbf{117}, 016602 (2016).
\bibitem{Ong} T. Liang, Q. Gibson, M. N. Ali, M. Liu, R. J. Cava, and N. P. Ong, Ultrahigh mobility and giant magnetoresistance in the Dirac semimetal Cd$_3$As$_2$, Nat. Mater. \textbf{14}, 280 (2015).
\bibitem{Zhao} Y. Zhao, H. Liu, C. Zhang et al., Anisotropic Fermi Surface and Quantum Limit Transport in High Mobility Three-Dimensional Dirac Semimetal Cd$_3$As$_2$, Phys. Rev. X \textbf{5}, 031037 (2015).
\bibitem{Chen} X. Huang, L. Zhao, Y. Long, P. Wang, D. Chen, Z. Yang, H. Liang, M. Xue, H.
Weng, Z. Fang, X. Dai, and G. Chen, Open Access Observation of the Chiral-Anomaly-Induced Negative Magnetoresistance in 3D Weyl Semimetal TaAs, Phys. Rev. X \textbf{5}, 031023 (2015).
\bibitem{Jia} C. Zhang, C. Guo, H. Lu, X. Zhang, Z. Yuan, Z. Lin, J. Wang, and S. Jia, Large magnetoresistance over an extended temperature regime in monophosphides of tantalum and niobium, Phys. Rev. B \textbf{92}, 041203 (2015).
\bibitem{Alidoust} M. Alidoust, K. Halterman, and A. A. Zyuzin, Superconductivity in
Type-II Weyl Semimetals, Phys. Rev. B \textbf{95}, 155124 (2017).
\bibitem{Li} D. Li, B. Rosenstein, B. Y. Shapiro, and I. Shapiro, Effect of the Type-I to Type-II Weyl Semimetal Topological Transition on Superconductivity, Phys. Rev. B \textbf{95}, 094513 (2017).
\bibitem{Brien} T. E. O'Brien, M. Diez, and C. W. J. Beenakker, Magnetic Breakdown and Klein Tunneling in a Type-II Weyl Semimetal, Phys.
Rev. Lett. \textbf{116}, 236401 (2016).
\bibitem{Zhang1} H. J. Zhang, C. X. Liu, X. L. Qi, X. Dai, Z. Fang, and S. C. Zhang, Topological insulators in Bi$_2$Se$_3$, Bi$_2$Te$_3$ and Sb$_2$Te$_3$ with a single Dirac cone on the surface, Nature Phys. \textbf{5}, 438-442 (2009).
\bibitem{Shen} Y. L. Chen, J. G. Analytis, J. H. Chu, Z. K. Liu, S. K. Mo, X. L. Qi, H. J. Zhang, D. H. Lu, X. Dai, Z. Fang, S. C. Zhang,
I. R. Fisher, Z. Hussain, and Z. X. Shen, Experimental Realization of a Three-Dimensional Topological Insulator, Bi$_2$Te$_3$, Science \textbf{325}, 178 (2009).
\bibitem{Ali} M. N. Ali, J. Xiong, S. Flynn, J. Tao, Q. D. Gibson, L. M. Schoop, T. Liang, N. Haldolaarachchige, M. Hirschberger, N. P. Ong,
and R. J. Cava, Large, Nonsaturating Magnetoresistance in WTe$_2$, Nature \textbf{514}, 205-208 (2014).
\bibitem{Balicas} D. Rhodes, R. Sch$\ddot{O}$nemann, N. Aryal, Q. Zhou, Q. R. Zhang, E. Kampert, Y. C. Chiu, Y. Lai, Y. Shimura, G. T. McCandless, J. Y. Chan, D. W. Paley, J. Lee; A. D. Finke, J. P. C. Ruff, S. Das, E. Manousakis, and L. Balicas, Bulk Fermi Surface of the Weyl Type-II Semimetallic Candidate $\gamma$-MoTe$_2$, Phys. Rev. B \textbf{96}, 165134 (2017).
\bibitem{PdTe2} Y. Liu, J. Z. Zhao, L. Yu, C. T. Lin, A. J. Liang, C. Hu, Y. Ding, Y. Xu, S. L. He, L. Zhao, G. D. Liu, X. L. Dong, J. Zhang, C. T. Chen, Z. Y. Xu, H. M. Weng, X. Dai, Z. Fang, and X. J. Zhou,Identification of Topological Surface State in PdTe$_2$ Superconductor by Angle-Resolved Photoemission Spectroscopy, Chin. Phys. Lett. \textbf{32}, 067303 (2015).
\bibitem{PtTe2} M. Yan, H. Huang, K. Zhang, E. Wang, W. Yao, K. Deng, G. H. Wan, H. Zhang, M. Arita, H. Yang, Z. Sun, H. Yao, Y. Wu, S. Fan, W. Duan, and S. Zhou, Lorentz-violating Type-II Dirac Fermions in Transition Metal Dichalcogenide PtTe$_2$, Nat. Commun. \textbf{8}, 257 (2017).
\bibitem{NiTe2} C. Q. Xu, B. Li, W. H. Jiao, W. Zhou, B. Qian, R. Sankar, N. D. Zhigadlo, Y. Qi, D. Qian, F. C. Chou, and X. Xu,
Topological type-II Dirac fermions approaching the Fermi level in a transition metal dichalcogenide NiTe$_2$, Chem. Mater. \textbf{30}, 4823 (2018).

\bibitem{TaIrTe4-1} K. Koepernik, D. Kasinathan, D. V. Efremov, Seunghyun Khim, Sergey Borisenko, Bernd B\"{u}chner, and Jeroen van den Brink, TaIrTe$_4$: A ternary type-II Weyl semimetal, Phys. Rev. B \textbf{93}, 201101 (2016).
\bibitem{MMTe4} J. W. Liu, H. Wang, C. Fang, L. Fu, and X. F. Qian, van der Waals Stacking-Induced Topological Phase Transition in Layered Ternary Transition Metal Chalcogenides, Nano Lett. \textbf{17}, 467 (2017).
\bibitem{TaIrTe4-2} Seunghyun Khim, K. Koepernik, D. V. Efremov, J. Klotz, T. Forster, J. Wosnitza, M. I. Sturza, S. Wurmehl, C. Hess, J. van den Brink,
and B. B\"{u}chner, Magnetotransport and de Haas-van Alphen measurements in the type-II Weyl semimetal TaIrTe$_4$, Phys. Rev. B 94, 165145 (2016).
\bibitem{NbIrTe4-1} R. Sch\"{o}nemann,  Y. C. Chiu,  W. K. Zheng, V. L. Quito, S. Sur, G. T. McCandless, J. Y. Chan, and L. Balicas,
Bulk Fermi surface of the Weyl type-II semimetallic candidate NbIrTe$_4$, Phys. Rev. B \textbf{99}, 195128 (2019).
\bibitem{NbIrTe4-2} W. Zhou, B. Li, C. Q. Xu, Maarten R. van Delft, Y. G. Chen, X. C. Fan, B. Qian, Nigel E. Hussey, and X. F. Xu, Nonsaturating Magnetoresistance and Nontrivial Band Topology of Type-II Weyl Semimetal NbIrTe$_4$, Adv. Electron. Mater. \textbf{5}, 1900250 (2019).
\bibitem{TaIrTe4-3} Y. Xing, Z. B. Shao, J. Ge, J. H. Wang, Z. W. Zhu, J. Liu, Y. Wang, Z. Y. Zhao, J. Q. Yan, D. Mandrus, B. H. Yan, X. J. Liu, M. H. Pan, and J. Wang, Surface superconductivity in the type-II Weyl semimetal TaIrTe$_4$, National Science Review, \textbf{nwz204} (2019).
\bibitem{TaPdTe16}W. H. Jiao, Z. T. Tang, Y. L. Sun, Y. Liu, Q. Tao, C. M. Feng, Y. W. Zeng, Z. A. Xu, and G. H. Cao, Superconductivity in a Layered Ta$_4$Pd$_3$Te$_{16}$ with PdTe$_2$ Chains, J. Am. Chem. Soc. \textbf{136}, 1284 (2014).
\bibitem{TaPdTe14}W. H. Jiao, L. P. He, Y. Liu, X. F. Xu, Y. K. Li, C. H. Zhang, N. Zhou, Z. A. Xu, S. Y. Li, and G. H. Cao, Superconductivity in Ta$_3$Pd$_3$Te$_{14}$ with quasi-one-dimensional PdTe$_2$ chains, Sci. Rep. \textbf{6}, 21628 (2016).
\bibitem{Bi4I4}G. Aut\`{e}s, A. Isaeva, L. Moreschini, J. C. Johannsen, A. Pisoni, R. Mori, W. T. Zhang, T. G. Filatova, A. N. Kuznetsov, L. Forr\'{o}, W. Broek, Y. Kim, K. S. Kim,
A. Lanzara, J. D. Denlinger, E. Rotenberg, A. Bostwick, M. Grioni, and O. V. Yazyev, A novel quasi-one-dimensional topological insulator in bismuth iodide $\beta$-Bi$_4$I$_4$, Nat. Mater. \textbf{15}, 154-158 (2016).
\bibitem{TaNiTe5}E. W. Liimatta and J. A. Ibers, Synthesis, Structures, and Conductivities of the New Layered
Compounds Ta$_3$Pd$_3$Te$_{14}$ and TaNiTe$_5$, J. Solid State Chem. \textbf{78}, 7-16 (1989).
\bibitem{Crystal} G. M. Sheldrick, A short history of SHELX, Acta Cryst. A\textbf{64}, 112-122 (2008).
\bibitem{SM} See Supplemental Material at *** for details on the independent atomic sites, thermal displacement parameters, the related bond lengths and bond angles of TaPdTe$_5$.
\bibitem{Wien2k} K. Schwarz, P. Blaha, and G. K. H. Madsen, Electronic structure calculations of solids using the WIEN2k package for material sciences, Comput. Phys. Commun. \textbf{147(1-2)}, 71 (2002).
\bibitem{GGA} Z. Wu and R. E. Cohen, More accurate generalized gradient approximation for solids, Phys. Rev. B \textbf{73}, 235116 (2006).
\bibitem{wannier1} A. A. Mostofi, J. R. Yates, G. Pizzi, Y.-S. Lee, I. Souza, D. Vanderbilt and N. Marzari, An updated version of wannier90: A Tool for Obtaining Maximally-Localised Wannier Functions, Comput. Phys. Commun. \textbf{185}, 2309 (2014).
\bibitem{wannier2} Q. S. Wu, S. N. Zhang, H. F. Song, M. Troyer, and A. A. Soluyanov, WannierTools: An open-source software package for novel topological materials, Comput. Phys. Commun. \textbf{224} 405-416 (2018).
\bibitem{NbNiTe5} E. W. Liimatta and J. A. Ibers, Synthesis, Structure, and Physical Properties of the New Layered
Ternary Chalcogenide NbNiTe$_5$, J. Solid State Chem. \textbf{71}, 384-389 (1987).
\bibitem{TaPtTe5} A. Mar and J. A. Ibers, Synthesis, Structure, and Physical Properties of the New Layered Ternary Telluride TaPtTe$_5$, J. Solid State Chem. \textbf{92}, 352-361 (1991).
\bibitem{NbPdTe5} E. W. Liimatta and J. A. Ibers, Synthesis, Structure, and Conductivity of the New Ternary
Chalcogenide NbPdTe$_5$, J. Solid State Chem. \textbf{77}, 141-147 (1988).
\bibitem{TaPdTe16-1} W. H. Jiao, Y. N. Huang, X. F. Xu, Y. K. Li, Y. Liu, Z. C. Wang, X. L. Xu, Y. X. Feng, C. M. Feng, and G. H. Cao, Normal-state properties of the quasi-one-dimensional superconductor Ta$_4$Pd$_3$Te$_{16}$, J. Phys.: Condens. Matter \textbf{31} 325601 (2019).
\bibitem{Core} G. A. Bain and J. F. Berry, Diamagnetic corrections and Pascal's constants, J. Chem. Edu. \textbf{85}, 532 (2008).
\bibitem{Wilson} K. G. Wilson, The renormalization group: Critical phenomena and the Kondo problem, Rev. Mod. Phys. \textbf{47}, 773 (1975).
\bibitem{Classical-hall} A. B. Pippard, Magnetoresistance in Metals (Cambridge: Cambridge University Press) (1989).
\bibitem{Cd3As2} T. Liang, Q. Gibson, M. N. Ali, M. Liu, R. J. Cava, and N. P. Ong, Ultrahigh Mobility and Giant Magnetoresistance in the Dirac
Semimetal Cd$_3$As$_2$. Nat. Mater. \textbf{14}, 280-284 (2015).
\bibitem{PdSn4} C. Q. Xu, W. Zhou, R. Sankar, X. Z. Xing, Z. X. Shi, Z. D. Han, B. Qian, J. H. Wang, Zengwei Zhu, J. L. Zhang, A. F. Bangura, N. E. Hussey, and Xiaofeng Xu, Enhanced electron correlations in the binary stannide PdSn$_4$: A homologue of the Dirac nodal arc semimetal PtSn$_4$, Phys. Rev. Mater. \textbf{1}, 064201 (2017).
\bibitem{Metal} K. Wang and C. Petrovic, Multiband Effects and Possible Dirac States in LaAgSb$_2$, Phys. Rev. B \textbf{86}, 155213 (2012).
\bibitem{YSb} O. Pavlosiuk, P. Swatek, and P. Wi\'{s}niewski, Giant magnetoresistance, three-dimensional Fermi surface and origin of resistivity plateau in YSb
semimetal, Sci. Rep. \textbf{6}, 38691 (2016).
\bibitem{Li} Y. Li, L. Li, J. Wang, T. Wang, X. Xu, C. Xi, C. Cao, and J. Dai, Resistivity plateau and negative magnetoresistance in the topological semimetal
TaSb$_2$, Phys. Rev. B \textbf{94}, 121115 (2016).
\bibitem{Turn-on} Y. L. Wang, L. R. Thoutam, Z. L. Xiao, J. Hu, S. Das, Z. Q. Mao, J. Wei, R. Divan, A. Luican-Mayer, G. W. Crabtree, and W. K. Kwok, Origin of the turn-on temperature behavior in WTe$_2$, Phys. Rev. B \textbf{92}, 180402 (2015).
\bibitem{ZrSiS} M. N. Ali, L. M. Schoop, C. Garg, J. M. Lippmann, E. Lara, B. Lotsch, and S. S. P. Parkin, Butterfly magnetoresistance, quasi-2D Dirac Fermi surface and topological phase transition in ZrSiS, Sci. Adv. \textbf{2}, e1601742 (2016).
\bibitem{LK} D. Shoenberg, Magnetic Oscillations in Metals (Cambridge Univ. Press, Cambridge, England, 1984).
\bibitem{Berry} G. P. Mikitik and Y. V. Sharlai, Manifestation of Berry's Phase in Metal Physics, Phys. Rev. Lett. \textbf{82}, 2147 (1999).
\bibitem{ZrSiSe} J. Hu, Z. Tang, J. Liu, X. Liu, Y. Zhu, D. Graf, K. Myhro, S. Tran, C. N. Lau, J. Wei, Z. Q. Mao, Evidence of Topological Nodal-Line Fermions in ZrSiSe and ZrSiTe, Phys. Rev. Lett. \textbf{117}, 016602 (2016).
\bibitem{Shift} J. Xiong, Y. K. Luo, YueHaw Khoo, S. Jia, R. J. Cava, and N. P. Ong, High-field Shubnikov-de Haas oscillations in the topological insulator Bi$_2$Te$_2$Se, Phys. Rev. B \textbf{94}, 165145 (2016).


\bibitem{Tl2Mo6Se6} Z. Song, B. Li, C. Xu, S. Wu, B. Qian, T. Chen, P. K. Biswas, X. Xu, and J. Sun, Pressure engineering of the dirac fermions in quasi-one-dimensional Tl$_2$Mo$_6$Se$_6$, J. Phys.: Condens. Matter \textbf{32}, 215402 (2020).
\bibitem{Cubic} Q. H. Liu and A. Zunger, Predicted Realization of Cubic Dirac Fermion in Quasi-One-Dimensional Transition-Metal Monochalcogenides, Phys. Rev. X \textbf{7}, 021019 (2017).
\bibitem{Fu2007} L. Fu and C. L. Kane, Topological insulators with inversion symmetry, Phys. Rev. B \textbf{76}, 045302 (2007).


\bibitem{zhang2009} H. Zhang, C.-X. Liu, X.-L. Qi, X. Dai, Z. Fang, S.-C. Zhang, Topological insulators in Bi$_2$Se$_3$, Bi$_2$Te$_3$ and Sb$_2$Te$_3$ with a single Dirac cone on the surface, Nat. Phys. \textbf{5} 438-442 (2009).

\end{thebibliography}
\end{document}